\documentclass[showpacs,twocolumn]{revtex4}
\usepackage{graphicx}
\usepackage{color}

\begin{document}

\title{Flavor Dependence of Meson Melting Temperature in Relativistic Potential Model}
\author{Shuzhe Shi}
\author{Xingyu Guo}
\author{Pengfei Zhuang}
\affiliation{Physics Department, Tsinghua University, Beijing 100084, China }
\date{\today}

\begin{abstract}
We study the melting temperature of heavy mesons in the hot medium of light quarks. By solving the covariant Schr\"odinger equations at finite temperature for mesons $D$, $\phi$ and $J/\psi$, we obtained the temperature dependence of their masses, binding energies and averaged sizes and found the flavor dependence of the melting temperature: $T_D \simeq T_\phi < T_{J/\psi}$. The sequential melting temperature can explain the difference in meson elliptic flow observed in heavy ion collisions.
\end{abstract}

\maketitle

From the lattice simulations of quantum chromodynamics (QCD) at finite temperature, there exists a deconfinement phase transition from hadron matter to quark matter at the critical temperature $T_c\simeq 170$ MeV~\cite{karsch}. Considering different binding energy of different hadrons, the melting temperature may depend on the flavor structure of hadrons. This flavor dependence of the deconfinement temperature is recently studied by lattice simulations~\cite{fodor,fodor2} and effective models~\cite{liao,huang,ratti,shi}. In the experiments of high energy nuclear collisions at the Relativistic Heavy Ion Collider (RHIC) and Large Hadron Collider (LHC), $J/\psi$, the bound state of charm quark $c$ and antiquark $\bar c$, is taken as a sensitive probe of the quark matter created in the early stage of the collisions~\cite{matsui}. The underlying reason is that the melting temperature of $J/\psi$ is much higher than the deconfinement temperature of light quarks and therefore it has the possibility to survive in the light quark matter and carries the information of the hot medium.

In this work, we study in the frame of relativistic potential model the flavor dependence of the melting temperature of those heavy mesons consisted of $s$ and $c$ quarks. While the quarkomium states ($c\bar c$ or $b\bar b$) can approximately be described in the non-relativistic potential model~\cite{satz}, one has to take into account the relativistic effects for light hadrons. The relativistic potential model has well been applied to the description of meson spectra in vacuum~\cite{crater1,crater4,crater11}. Taking Pauli reduction and scale transformation~\cite{long}, the two-body Dirac equation is effectively expressed as a group of covariant Schr\"odinger equations and used to calculate the wave functions of $q\bar q$ bound states at $T=0$~\cite{crater2}. We extended this model to $c\bar c$ bound states at finite temperature and found that the relativistic correction to the $c\bar c$ melting temperature is about $10\%$~\cite{guo}. The power of the relativistic potential model at finite temperature is not only the correction to the quarkonium states, but also a reasonable description to those hadrons consisted of $s$ and $c$ quarks. In this paper we calculate the melting temperature of $D$, $\phi$ and $J/\psi$ mesons in the hot medium of light quarks and see if there exists a sequential deconfinement phase transition from light to heavy quarks.

In order to describe a general bound state $q_1\bar q_2$ with $q_1, q_2=u,d,s,c$, we consider the following Schr\"odinger equations~\cite{crater2} for the radial motion of the $q_1\bar q_2$ state relative to the center of mass. The radial wave functions of the spin singlet $u_0$ and one of the spin triplet $u_1^0$ with quantum numbers $n^{2s+1}l_j=n^1l_l$ and $n^3l_l$ are controlled by the two coupled equations,
\begin{widetext}
\begin{eqnarray}
\label{wave1}
&& \left[-\frac{d^2}{dr^2}+\frac{j(j+1)}{r^2}+2m_wB+B^2+2\epsilon_wA-A^2+\Phi_D-3\Phi_{SS}\right]u_0+2\sqrt{j(j+1)}\left(\Phi_{SOD}-\Phi_{SOX}\right)u_1^0=b^2u_{0}, \nonumber\\
&& \left[-\frac{d^2}{dr^2}+\frac{j(j+1)}{r^2}+2m_wB+B^2+2\epsilon_wA-A^2+\Phi_D-2\Phi_{SO}+\Phi_{SS}+2\Phi_T-2\Phi_{SOT}\right]u_1^0\nonumber\\
&& +2\sqrt{j(j+1)}\left(\Phi_{SOD}+\Phi_{SOX}\right)u_0=b^2u_1^0,
\end{eqnarray}
and the other two states $u_1^+$ and $u_1^-$ of the triplet with quantum numbers  $n^3l_{l+1}$ and $n^3l_{l-1}$ are characterized by
\begin{eqnarray}
\label{wave2}
&& \left[-\frac{d^2}{dr^2}+\frac{j(j-1)}{r^2}+2m_wB+B^2+2\epsilon_wA-A^2+\Phi_D+2(j-1)\Phi_{SO}+\Phi_{SS}+\frac{2(j-1)}{2j+1}\left(\Phi_{SOT}-\Phi_T\right)\right]u_1^+\nonumber\\
&& +\frac{2\sqrt{j(j+1)}}{2j+1}\left(3\Phi_T-2(j+2)\Phi_{SOT}\right)u_1^- =b^2u_1^+,  \nonumber\\
&& \left[-\frac{d^2}{dr^2}+\frac{(j+1)(j+2)}{r^{2}}+2m_wB+B^2+2\epsilon_wA-A^2+\Phi_D-2(j+2)\Phi_{SO}+\Phi_{SS}+\frac{2(j+2)}{2j+1}\left(\Phi_{SOT}-\Phi_T\right)\right]u_1^-\nonumber\\
&& +\frac{2\sqrt{j(j+1)}}{2j+1}\left(3\Phi_T+2(j-1)\Phi_{SOT}\right)u_1^+ =b^2u_1^-
\end{eqnarray}
with the energy eigenvalue
\begin{equation}
\label{eigrn}
b^2={1\over 4}\left[m_m^2-2\left(m_{q1}^2+m_{q2}^2\right)+\left(m_{q1}^2-m_{q2}^2\right)^2/m_m^2\right],
\end{equation}
\end{widetext}
where $n$ is the principal quantum number, $l$, $s$ and $j$ are the orbital, spin and total angular momentum numbers, $m_m$, $m_{q1}$ and $m_{q2}$ are the meson and quark masses, the other two mass parameters $m_w$ and $\epsilon_w$ are defined as $m_w=m_{q1}m_{q2}/m_m$ and $\epsilon_w=\left(m_m^2-m_{q1}^2-m_{q2}^2\right)/(2m_m)$, and the explicit expressions for the Darwin term, spin-spin and spin-orbit couplings and tensor terms $\Phi_D, \Phi_{SS}, \Phi_{SO}, \Phi_T, \Phi_{SOT}, \Phi_{SOD}$ and $\Phi_{SOX}$ introduced in the dynamical equations are given in Ref.~\cite{crater2}. Note that for $q_1=q_2$ the tensor terms $\Phi_{SOD}$ and $\Phi_{SOX}$ disappear and the two wave equations (\ref{wave1}) become decoupled. For the $S$-wave of spin singlet $^1S_0$ with $j=0$, such as $D$ and $D_s$ mesons, the mixing terms vanish too.

In the Schr\"odinger equations (\ref{wave1}) and (\ref{wave2}) the central potential between the quark $q_1$ and antiquark $\bar q_2$ has been separated into two parts~\cite{crater2},
\begin{equation}
\label{vr}
V(r)=A(r) + B(r),
\end{equation}
$A$ and $B$ control respectively the behavior of the potential at short and long distance. In vacuum we take the Cornell potential, including a Coulomb-like part which dominates the wave
functions around $r=0$ and a linear part which leads to the quark confinement,
\begin{eqnarray}
\label{vacuum}
A(r)&=&-{\alpha\over r},\nonumber\\
B(r)&=&\sigma r.
\end{eqnarray}
Since the interaction between quarks is a color interaction, the coupling constants $\alpha$ and $\sigma$ are independent of the flavor structure of quarks $q_1$ and $\bar q_2$. Different from the previous works~\cite{crater1,crater2,guo} where the parameters in the Schr\"odinger equations are fixed by fitting the meson masses in vacuum, we determine, like Ref.\cite{satz}, the parameters by considering both the meson masses in vacuum and the lattice calculated quark potential~\cite{kaczmarek,datta,digal} at finite temperature. We take $\alpha=\pi/12$ and $\sigma=0.19$ GeV$^2$ which fit the lattice data well and the quark masses $m_c$=1.28 GeV, $m_s$=0.08 GeV and $m_u=m_d$=0.005 GeV which lead to the meson masses shown in Table \ref{table1}.
\begin{table}[h!t]
\caption{Meson masses in vacuum and the comparison with the experimental data~\cite{exp}.}
\label{table1}
\begin{tabular}{|c c|c c|}
\hline
Meson		&$n^{2s+1}l_j$			&Exp.(GeV)	&Theo.(GeV)	\\
\hline
$\phi:s\bar s$	    &$1^{3}S_1+1^{3}D_1$		&1.019	&1.096	\\
$D:c\bar u$		    &$1^{1}S_0$			        &1.865	&1.929	\\
$D^*:c\bar u	$    &$1^{3}S_1+1^{3}D_1$		&2.010	&1.989	\\
$D_s:c\bar s $   &$1^{1}S_0$	        		&1.968	&1.978	\\
$D^*_s:c\bar s$	    &$1^{3}S_1+1^{3}D_1$		&2.112	&2.037	\\
$J/\psi:c\bar c$    &$1^{3}S_1+1^{3}D_1$		&3.097	&3.045	\\
$~~\psi':c\bar c$	&$2^{3}S_1+1^{3}D_1$		&3.686	&3.609	\\
$~~\chi_1:c\bar c$	&$1^{3}P_1$		        	&3.511	&3.395	\\
\hline
\end{tabular}
\end{table}

At finite temperature the free energy $F$ of a pair of heavy quarks is calculated by lattice simulations~\cite{kaczmarek,digal,blaschke}. It is the potential in the limit of slow meson dissociation in the medium. In this case there is enough time for the meson to exchange heat with the medium. However, in the limit of rapid dissociation, there is no heat exchange between the meson and the medium, and the potential is the internal
energy $U$ which is related to $F$ through the thermodynamic relation $U=F-T\partial F/\partial T$. In general case the quark potential is in between the two limits. When temperature $T$ vanishes, there is no more difference between $F$ and $U$, and we come back to the Cornell potential (\ref{vacuum}). While different quark potential at finite temperature will change the meson wave functions, the flavor dependence of the meson melting temperature $T_m$, especially the order of $T_d$s is not sensitive to the choice of $V$. In the following calculations at finite temperature we take the limit $V=F$ as an example and will discuss the difference in the other limit $V=U$. Considering the Debye screening in the medium, the potential
$V(r,T)=A(r,T)+B(r,T)=F(r,T)$ can be written as~\cite{digal,satz}
\begin{eqnarray}
\label{vf}
A(r,T)&=&-{\alpha\over r} e^{-\mu r},\\
B(r,T)&=&{\sigma\over \mu}\left[{\Gamma\left({1\over 4}\right)\over
2^{3\over 2}\Gamma\left({3\over 4}\right)} -{\sqrt{\mu r}\over
2^{3\over 4}\Gamma\left({3\over 4}\right)} K_{1\over 4}\left(\mu^2
r^2\right) \right]-\alpha\mu,\nonumber
\end{eqnarray}
where $\Gamma$ is the Gamma function, $K$ is the modified Bessel
function of the second kind, and the temperature dependent parameter
$\mu(T)$, namely the screening mass or the inverse screening radius,
can be extracted from fitting the lattice simulated free
energy~\cite{digal,kaczmarek}.

Using the inverse power method~\cite{crater3} to solve the Schr\"odinger equations (\ref{wave1}) and (\ref{wave2}), we obtain the meson radial wave function
\begin{equation}
\label{radial}
\psi(r,T)={u(r,T)\over r}
\end{equation}
and the meson mass $m_m(T)$ through the energy eigenvalue $b^2(T)$. From the known temperature dependence of the meson mass and wave function, we derive in turn the relativistic meson binding energy~\cite{guo}
\begin{equation}
\label{binding}
\epsilon(T) = V(\infty,T)+\sqrt{V^2(\infty,T)+(m_{q1}+m_{q2})^2}-m_m(T)
\end{equation}
and the averaged meson size, namely the distance between the quarks $q_1$ and $\bar q_2$,
\begin{equation}
\label{r}
\langle r\rangle (T)={\int dr r^3\left|\psi(r,T)\right|^2
\over \int dr r^2\left|\psi(r,T)\right|^2}.
\end{equation}

\begin{figure}[!h]
\begin{center}
\includegraphics[width=0.55\textwidth]{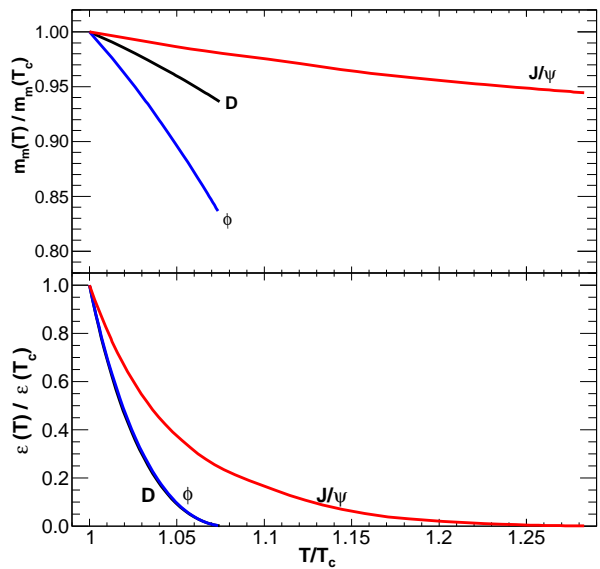}
\caption{(color online) The scaled meson mass $m_m(T)/m_m(T_c)$ (upper panel) and scaled meson binding energy $\epsilon(T)/\epsilon(T_c)$ (lower panel) as functions of scaled temperature $T/T_c$ for $D$, $\phi$ and $J/\psi$. $T_c$ is the deconfinement temperature of light quarks, and the model calculations are with quark potential $V=F$. }
\label{fig1}
\end{center}
\end{figure}

Since the mass change for a heavy quark system is expected to be weak at low temperature, it is normally neglected in model calculations. However, from the calculations with QCD sum rules~\cite{sumrule1,sumrule2} and QCD second-order Stark effect~\cite{stark}, the $J/\psi$ mass is remarkably changed in a static hot medium. In the
region above and close to the deconfinement temperature $T_c$ of light quarks, there is a strong change in the mass of $J/\psi$. For instance, at
temperature $T/T_c= 1.1$ the mass shift $\Delta m_{J/\psi}=m_{J/\psi}(T)-m_{J/\psi}(T_c)$ can reach 100 MeV~\cite{stark}, which is
already comparable with the mass change for light hadrons~\cite{light}. From our calculation in the frame of relativistic potential model at finite temperature, shown in the upper panel of Fig.\ref{fig1}, the hot medium effect on the meson mass is remarkable too. The maximum mass reduction is about $5\%$ for $J/\psi$ and $D$ and reaches $20\%$ for $\phi$.

While the change in the meson mass is smooth, the binding energy $\epsilon(T)$ drops down very fast due to the rapidly decreasing potential $V(\infty,T)$ at infinite distance, see the lattice calculations~\cite{digal,satz}. This means that the fast melting of heavy mesons in hot medium is not due to their mass change but from the strong Debye screening which changes dramatically the potential between the two quarks. From the definition of the meson melting temperature $T_m$,
\begin{equation}
\label{melt}
\epsilon(T_m)=0,
\end{equation}
the meson can survive in the light quark matter in the temperature region $T_c<T<T_m$. Note that the binding energy is controlled by the constant quark masses $m_{q1}$ and $m_{q2}$ and the temperature dependent meson mass $m_m(T)$ and the quark potential $V(\infty, T)$, see equation (\ref{binding}), the melting temperature is determined by their competitions. The values of $T_m$ extracted from Fig.\ref{fig1} are shown in Table \ref{table2}. Since the potential $V=U$ is much stronger than the potential $V=F$, the temperature needed to melt the meson is much higher in $V=U$ than that in $V=F$. Considering the fact that the real quark potential is in between $F$ and $U$, the melting temperature is in a wide region, $1.1\lesssim T_D/T_c \simeq T_\phi/T_c\lesssim 1.8$ and $1.3\lesssim T_{J/\psi}/T_c\lesssim 2.5$.
\vspace{-0.5cm}
\begin{table}[h!t]
\caption{Meson melting temperature $T_m$ for $D$, $\phi$ and $J/\psi$ in the two limits of quark potential $V=F$ and $V=U$.}
\label{table2}
\begin{tabular}{|c|c|c|}
\hline
meson&$T_m/T_c$(V=F)&$T_m/T_c$(V=U)\\
\hline
$D$&1.08&1.81\\
$\phi$&1.08&1.77\\
$J/\psi$&1.28&2.51\\
\hline
\end{tabular}\label{td}
\end{table}

\begin{figure}[!h]
\begin{center}
\includegraphics[width=0.55\textwidth]{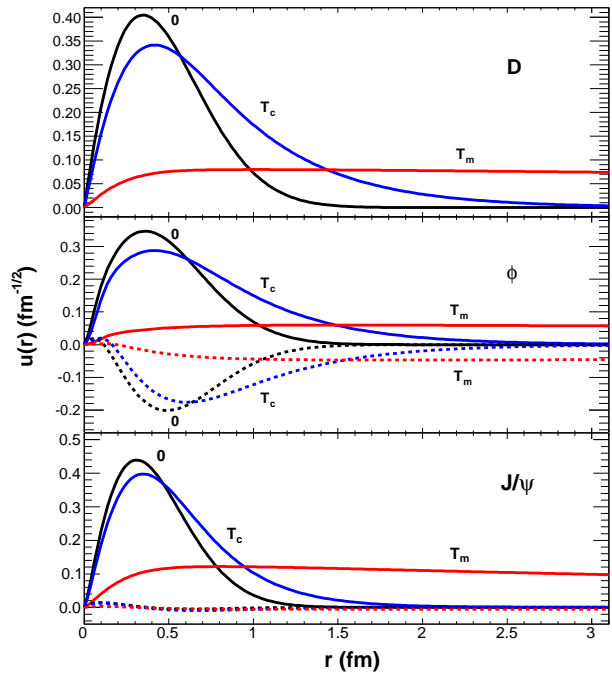}
\caption{(color online) The radial wave function $u(r)$ as a function of the distance $r$ between the two quarks for mesons $D$ (upper panel), $\phi$ (middle panel) and $J/\psi$ (lower panel) at three temperatures $T=0, T_c$ and $T_m$. The solid and dashed lines are respectively for the $S$ and $D$ waves. $T_c$ is the deconfinement temperature of light quarks, $T_m$ is the heavy meson melting temperature, and the model calculations are with quark potential $V=F$. }
\label{fig2}
\end{center}
\end{figure}

The radial wave functions for mesons $D$, $\phi$ and $J/\psi$ are shown in Fig.\ref{fig2}. In vacuum with $T=0$, the wave functions are mainly distributed in a narrow region of $r<1$ fm and the peaks are located at $r\sim 0.3$ fm. This means quark confinement in vacuum. With increasing temperature of the system, the wave functions expand continuously. At the deconfinement temperature $T_c$ of the light quarks, while the wave functions shift outside a little, the distribution is still similar to the one in vacuum. This indicates that the heavy mesons can survive in the soup of light quarks. The change from $T_c$ to the meson melting temperature $T_m$ is however dramatic, and the wave functions expand rapidly. This means the collapse of the heavy meson systems. For vector mesons $\phi$ and $J/\psi$, their wave functions contain two components, the $S$ and $D$ waves, shown as solid and dashed lines in Fig.\ref{fig2}. Since charm quark is much heavier than strange quark, the relative rotation between the $c$ and $\bar c$ should be much weaker than that between the $s$ and $\bar s$. As a consequence, the $D$-wave of $J/\psi$ can be neglected at any temperature, but the $D$- and $S$-waves for $\phi$ are almost equally important.

\begin{figure}[!h]
\begin{center}
\includegraphics[width=0.55\textwidth]{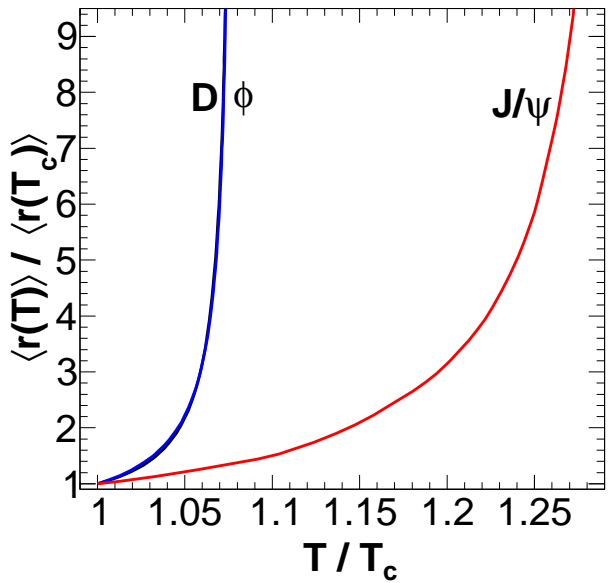}
\caption{(color online) The scaled average size $\langle r\rangle(T)/\langle r\rangle(T_c)$ as a function of the scaled temperature $T/T_c$ for mesons $D$, $\phi$ and $J/\psi$. $T_c$ is the deconfinement temperature of light quarks, and the model calculations are with quark potential $V=F$. }
\label{fig3}
\end{center}
\end{figure}

The melting temperature $T_m$ can also be defined through the infinite size of the meson,
\begin{equation}
\label{size}
\langle r\rangle(T_m)\to\infty,
\end{equation}
which is equivalent to the definition of zero binding energy. The scaled average size $\langle r\rangle (T)/\langle r\rangle(T_c)$ as a function of scaled temperature $T/T_c$ is shown in Fig.\ref{fig3}. When the temperature approaches to the melting temperature, the meson size increases dramatically, and the meson collapse process is very fast.
\vspace{-0.5cm}
\begin{figure}[!h]
\begin{center}
\includegraphics[width=0.55\textwidth]{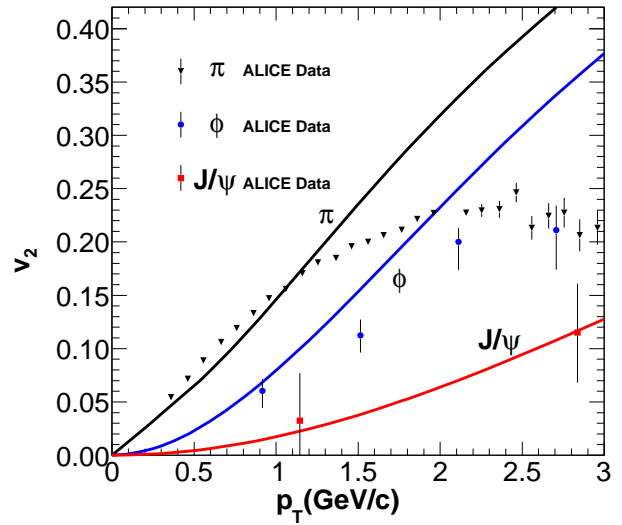}
\caption{(color online) The meson elliptic flow $v_2$ as a function of its transverse momentum $p_t$ for $\pi$, $\phi$ and $J/\psi$ in Pb+Pb collisions at $\sqrt {s_{NN}}=2.76$ TeV. The experimental data are from \cite{v2phi} for $\pi$ and $\phi$ at centrality bin $20\%-60\%$ and rapidity bin $|\eta|<0.8$ and \cite{v2jpsi} for $J/\psi$ at bins of  $40\%-50\%$ and $2.5<\eta<4.0$, and the model calculations are at impact parameter $b=10.2$ fm and with quark potential $V=F$. }
\label{fig4}
\end{center}
\end{figure}

To see possible effect of the sequential hadron melting temperature on the finally observed distributions in high energy heavy ion collisions, we estimate now the meson elliptic flow $v_2$ in the frame of hydrodynamics~\cite{heinz,hirano}. $v_2$ is created in the initial stage of the collisions and develops in the hot medium, it is therefore sensitive to the hadron melting temperature. At LHC energy the initial temperature of the colliding system is so high, the $u, d, s$ and $c$ quarks are all deconfined in the early stage of the hydrodynamic evolution of the fireball. With the expansion of the system, the temperature goes down and mesons are formed at the melting temperature $T_m$. Taking the ideal hydrodynamics $\partial_\mu T^{\mu\nu}=0$ with $T^{\mu\nu}$ being the energy-momentum tensor and the equation of state with a first order phase transition between partons and hadrons~\cite{heinz}, we obtain the meson momentum distribution
\begin{equation}
\label{hydro}
{dN\over d{\bf p}_tdy} = {1\over (2\pi)^3}\int d\sigma_\mu p^\mu f_m
\end{equation}
where $\sigma_\mu$ is the meson formation hypersurface determined by the melting temperature $T({\bf x},t)=T_m$, $f_m=1/\left(e^{p_u u^\mu/T_m}-1\right)$ is the thermalized meson distribution at $T_m$, and the local temperature $T$ and fluid velocity $u_\mu$ are from the solution of the hydrodynamics. We have neglected here the meson interactions in the hadron phase. The meson elliptic flow is defined as
\begin{equation}
\label{v2}
v_2={\int d\varphi\ dN/d\varphi \cos(2\varphi)\over \int\ d\varphi dN/d\varphi}
\end{equation}
with $\varphi$ being the angle between the short axis of the ellipse and the transverse momentum ${\bf p}_t$. For Pb+Pb collisions at $\sqrt{s_{NN}}=2.76$ TeV  and impact parameter $b=10.2$ fm, we calculated the elliptic flow for $\pi$ and $\phi$ in central rapidity and $J/\psi$ in forward rapidity, the result and the comparison with the experimental data are shown in Fig.\ref{fig4}. From the data there is the relation $v_2^\pi > v_2^\phi > v_2^{J/\psi}$. This can be understood from the flavor dependence of the melting temperature $T_\pi(=T_c)<T_\phi <T_{J/\psi}$: A high melting temperature means an early hadronization of the corresponding quarks in heavy ion collisions, and therefore these quarks do not have enough time to develop the elliptic flow. The model calculations agree reasonably well with the data at low $p_t$ where the hydrodynamics works.

In summary, we investigated the melting temperature of the mesons consisted of $s$ and $c$ quarks in the hot medium of light quarks. In the relativistic potential model, we solved the covariant Schr\"odinger equations for $D$, $\phi$ and $J/\psi$ at finite temperature, with the help of the central potential extracted from the lattice simulations. We obtained the meson binding energy and average size which determine the melting temperature. We found a sequential melting temperature $T_D \simeq T_\phi < T_{J/\psi}$, which can be used to explain the difference in meson elliptic flows observed at LHC.

\appendix {\bf Acknowledgement:} The work is supported by the NSFC under grant No. 11079024 and the MOST under grant No. 2013CB922000.

\bibliographystyle{unsrt}

\end{document}